\documentclass[12pt]{article}
\usepackage[dvips]{graphicx}

\newlength{\extraspace}
\newlength{\extraspaces}
\newcommand{\be}{\begin{equation}
\addtolength{\abovedisplayskip}{\extraspaces}
\addtolength{\belowdisplayskip}{\extraspaces}
\addtolength{\abovedisplayshortskip}{\extraspace}
\addtolength{\belowdisplayshortskip}{\extraspace}}
\newcommand{\ee}{\end{equation}}

\newcommand{\ba}{\begin{eqnarray}
\addtolength{\abovedisplayskip}{\extraspaces}
\addtolength{\belowdisplayskip}{\extraspaces}
\addtolength{\abovedisplayshortskip}{\extraspace}
\addtolength{\belowdisplayshortskip}{\extraspace}}
\newcommand{\ea}{\end{eqnarray}}

\title{ A generalization of Abel Inversion to non axisymmetric density distribution}
\author{\bf{P. Tomassini and A. Giulietti}\\
\\
Istituto di Fisica Atomica e Molecolare \\ CNR Area della Ricerca
di Pisa\\ Via G. Moruzzi, 56124 Pisa, Italy\\ E-mail:
tomassini@ifam.pi.cnr.it
\\
}

\begin{document}

\maketitle

\begin{abstract}
Abel Inversion is currently used in laser-plasma studies in order
to estimate the electronic density $n_e$ from the phase-shift map
$\delta \phi$ obtained via interferometry. The main limitation of
the Abel method is due to the assumption of axial symmetry of the
electronic density, which is often hardly fulfilled. In this paper
we  present an improvement to the Abel inversion technique in
which the axial symmetry condition is relaxed by means of
 a truncated Legendre Polinomial expansion in the azimutal angle.
With the help of simulated interferograms, we will show that the
generalized Abel inversion generates accurate densities maps when
applied to non axisymmetric density sources.
\end{abstract}
\par

Abel Inversion is widely used in many context and, in laser-plasma
studies it leads to a 2D electronic density map $n_e$
reconstruction from phase-shift maps $\delta \phi$ recorded using
interferometry \cite{Nomarski}\cite{Gizzi}. Once the phase map
$\delta \phi(z,x)$ has been extracted from the interferogram via
standard FFT techniques \cite{Nugent} or with a more sophisticated
Wavelet-Based method \cite{Tomassini}, the best symmetry axis
$z_0$ should be defined and two half-space phase maps
 \ba \label{cut}
\delta\phi^+(\zeta,x) & = & \delta\phi(z-z_0,x) \, \, z>z_0
\nonumber\\
 \delta\phi^-(\zeta,x) & = & \delta\phi(z_0-z,x) \,
\, z<z_0 \ea \noindent are introduced. By assuming axial symmetry
around the laser-beam propagation axis $x$,
$\delta\phi^+=\delta\phi^-$ so the electronic density map
$n_e(r,x)$ is computed as
\be
\label{abel}
 n_e(r,x) = -n_c{\lambda_p\over \pi^2} \int_r^{\infty}d\zeta
{1\over \sqrt{\zeta^2-r^2}}{\partial \over \partial \zeta}
\delta\phi^{\pm}(\zeta,x) \ee \noindent where  $\lambda_p$ is the
probe wavelength and $n_c = \pi m (c/(e \lambda_p))^2$ is the
critical density at $\lambda_p$.

Though laser plasmas show approximately axial symmetry in general,
significant deviations from the symmetry may occur. In these
latter cases the Abel Inversion applied to an artificial profile
obtained by symmetrization of either $\delta\phi^+$ or
$\delta\phi^-$, can lead to misleading reconstruction of the
density distribution. Alternately, to consider the "mean" phase
distribution $\delta\phi_s = 1/2 (\delta\phi^++\delta\phi^-)$ can
also induce large errors.

In $1981$ Yasutomo {\it et al} \cite{Yasutomo} presented a
generalization of Abel inversion based on the assumption that the
$3D$ density distribution $n(r,x,\theta) $ can be factorized as
the product of a isotropic density $\tilde{n}_0(r)$ and a
corrective term $g(r \sin\theta)$, being $\theta$ the azimutal
angle.

In this paper we introduce a generalization of the Abel inversion
algorithm to be applied to moderately asymmetric interferograms.
Unlike Yasutomo, we base our algorithm on a Legendre polinomial
expansion of $n(r,x,\theta) $ in the angular variable alone.
  We will
show that such an extension of the Abel method allows accurate
reconstructions of the density distribution in some simulated non
symmetric cases.

The basic geometry of the phase-shift acquisition via
interferometry is shown in Fig. \ref{fig:figura1}, in which a
plane parallel to the laser propagation axis ($x$-direction), Fig.
\ref{fig:figura1} a), and a plane perpendicular to that axis, Fig.
\ref{fig:figura1} b), are shown, respectively.

Let us point out that any departure from the mirror symmetry
respect to the plane perpendicular to the probe axis (the $x-z$
plane in Fig. \ref{fig:figura1}) cannot be taken into account
because of the line integral in the acquisition step. Consequently
we can assume that such a mirror symmetry is satisfied.

 The phase-shift
$\delta\phi(z,x)$ detected in the $(z,x)$ position on the
interferogram is then linked to the electronic density
$n_e(r,\theta,x)$ as: \ba \label{deltaphias}
 \delta\phi(z,x) &=& -{\pi\over \lambda_p}
\int_{-\infty}^{\infty}{n(z,y,x)\over n_c}dy \nonumber\\ &=&
-{2\pi\over \lambda_p}
\int_{|z-z_0|}^{\infty}{r\over\sqrt{r^2-(z-z_0)^2}}{n(r,\theta(r),x)\over
n_c} dr \, ,
 \ea
\noindent where we moved to the cylindrical coordinates
$(r,\theta)$, taking $x$ as symmetry axis.

 We now make the physically justified assumption that the angular
dependence of $n_e(r,\theta,x)$ is everywhere "well behaved" (no
abrupt changes occur) so that $n_e$ can be developed as a
truncated series of orthonormal  Legendre Polinomials
$P_l(cos(\theta))$:
\be
\label{dev} n_e(r,\theta,x) =
\sum_{l=0}^Ln_l(r,x)P_l(cos(\theta))\, . \ee To find the
appropriate value of $L$, let us simply observe that for each x
from $\delta\phi^+(\zeta)$ and $\delta\phi^-(\zeta)$ we can build
up two linearly independent sequences $$\delta\phi_s \equiv
{1\over 2}\left(\delta\phi^+(\zeta)+\delta\phi^-(\zeta)\right) $$
 $$\delta\phi_a \equiv {1\over
2}\left(\delta\phi^+(\zeta)-\delta\phi^-(\zeta)\right) $$
 so that for each $x$ and $r$ we have {\it two} independent degrees of freedom
 which could be linked to the angular dependence and this leads to
 $L = 1$. The truncation of the series of Eq. \ref{dev} up to $l=1$ is
 straightforward. Since $P_0(x) = 1$ and $P_1(x) = x$, we have
\be\label{poly} n(r,\theta,x) = n_0(r,x)+n_1(r,x)\, cos(\theta)
\ee \noindent and so the phase-shift map is computed as: \ba
\label{phase3}
 \delta\phi(z,x)= -{2\pi\over
\lambda_p}\int_{|z-z_0|}^{\infty}{r\over\sqrt{r^2-(z-z_0)^2}}{1\over
n_c} \left\{n_0(r,x)+n_1(r,x){(z-z_0)\over r} \right\}dr \, .
 \ea
\noindent Defining $\nu_1(r,x)\equiv n_1(r,x)/r$ and extracting
the symmetric and antisymmetric components of $\delta\phi$ in Eq.
\ref{phase3}, we obtain
 \ba
 \label{phase4}
  \delta\phi_s(
  \zeta,x) &=& -{\pi\over
\lambda_p}
\int_{\zeta}^{\infty}{r\over\sqrt{r^2-\zeta^2}}{n_0(r,x)\over
n_c}dr \nonumber\\
 \delta\phi_a(\zeta,x) &=& -{\pi \zeta \over
\lambda_p}
\int_{\zeta}^{\infty}{r\over\sqrt{r^2-\zeta^2}}{\nu_1(r,x)\over
n_c}dr \ea\, \noindent and we can finally invert Eq. \ref{phase4}
obtaining the coefficients of the generalized Abel Inversion:
 \ba \label{GenAbel} n_0(r,x) &=& -n_c {\lambda_p\over
\pi^2} \int_r^{\infty}d\zeta {1\over \sqrt{\zeta^2-r^2}}{\partial
\over\partial \zeta} \delta\phi_s(\zeta,x) \nonumber\\
 n_1(r,x) &=& -n_c {\lambda_p\over
\pi^2} r \int_r^{\infty}d\zeta {1\over
\sqrt{\zeta^2-r^2}}{\partial \over\partial \zeta}
\left({\delta\phi_a(\zeta,x)\over \zeta}\right)\, . \ea

The application of the generalized Abel inversion (equations (
\ref{poly}), (\ref{GenAbel})) is straightforward and very
effective. In order to prove this, the new algorithm will be
tested with two sample interferograms. Both of them have been
obtained numerically from {\it a priori} known density
distributions, with which the reconstructed distributions can be
compared.

Let us firstly consider the sample interferogram of Fig.
\ref{fig:P1} b), obtained from the 3D density distribution of the
form $n^{True}(r,\theta,x)= n_0(r,x)\,
P_0(\cos(\theta))+n_1(r,x)\, P_1(\cos(\theta))+n_2(r,x)\,
P_2(\cos(\theta))$, with $n_i$ gaussian shaped in the radial
direction $r$ and exponentially decreasing in the longitudinal
direction $x$. The simulation is performed assuming a $1 \mu m$
wavelength probe and a maximum density of $n_i$ in $n_c$ units as
$n_0^{max} = 0.1$, $n_1^{max} = 0.05$, $n_2^{max} = 0.025$.

Once the phase shift $\delta\phi$ has been extracted from the
interferogram, an automatic procedure to optimize the position of
the global symmetry axis ($z_0$) has been applied and the two half
phase maps $\delta\phi^{+}$ and $\delta\phi^{-}$ have been
constructed. Next, the electronic densities $n^{+}(r,x)$,
$n^{-}(r,x)$, $n^{mean}(r,x)$ are computed {\it via} standard Abel
Inversion applied to $\delta\phi^+$, $\delta\phi^-$,
$\delta\phi_{s}= 1/2(\delta\phi^{+}+\delta\phi^{-})$,
respectively. Finally, the generalized Abel inversion (Eqq.
\ref{poly}, \ref{GenAbel}) is applied to both $\delta\phi^+$ and
$\delta\phi^-$ producing $n^{gener}(r,\theta,x)$.

In Fig. \ref{fig:P1} projections onto the $z-x$ plane of the
simulated density (a), the standard Abel inversion of  the
symmetryzed map $\delta\phi_s$ (c) and of the generalized Abel
inversion of $\delta\phi^+$ and $\delta\phi^-$ (d), are shown. As
it is clear, $n^{mean}$ shape differs considerably from the one of
$n^{True}$, while $n^{Gener}$ well match the true density map. For
a more quantitative comparison, we have reproduced   in Fig.
\ref{fig:P2} b) line-outs of the projection of the true density
map $n^{True}$ and of $n^{+}$ and $n^{-}$ at $x = 10 \mu m$ from
the simulated target. As it is evident, not only  none of them
reasonably reproduces the true density contour but their shapes
also differ very much. As a result, standard Abel Inversion is not
applicable in this case in order to produce a (reasonably)
accurate density map. In Fig. \ref{fig:P2}  a) the line-outs of
the projection of the generalized Abel inversion $n^{gener}$ and
the standard inversion of $\delta\phi_s$ are confronted with the
true density contour (dotted line) and $n_{01}\equiv n_0 \,
P_0+n_1\, P_1$ (dashed line). Standard Abel inversion applied to
the mean phase-shift map still fails in reproducing a reasonable
density map, while generalized Abel inversion gives us a contour
which is everywhere well overlapped to the true one. Now, as a
result, we can claim that with the use of the generalized Abel
Inversion a good estimation of the simulated density map is
achieved.

Now, let us  test the new algorithm in a physical condition which
is often experimentally observed: a density with an axially
symmetric background to which it is added an axially symmetric
channel whose symmetry axis is not aligned with the one of  the
background (see Fig. \ref{fig:C1} a)). As before, the maximum
electronic density is well below the critical density
$n^{True}_{max}= 0.1\, n_c$ and the probe wavelength is $1\mu m$.

As in the previous example we apply the standard and the
generalized Abel inversions to the phase-shift maps extracted by
the interferogram reproduced in Fig. \ref{fig:C1} b). In Figg.
\ref{fig:C1} c) and d),  the best output of the standard Abel
inversion (the one obtained with $\delta\phi_+$) and of the
generalized Abel inversion are shown, respectively. As in the
previous example, standard Abel inversion produces a poorly
accurate density map. A more quantitative analysis can be
performed with the help of line-outs reported in Fig.
\ref{fig:C2}, which shows that standard Abel inversion results
should be rejected. On the contrary,
  generalized Abel inversion produces reasonably good results
in almost all the density map but a thin band near the best
symmetry axis, where the dependence of $n^{True}$ on
$\cos(\theta)$ is much more  complex then linear.

We face now with a noisy phase map in order to compare the noise
content in the standard and in the generalized Abel inversions.
Here we will focus only on Gaussian and uncorrelated (white)
noise, which is added to the phase-shift map of the interferogram
reproduced in Fig.  \ref{fig:C1} b) (see Fig. \ref{fig:noise}
(a)). To visualize the noise which is propagated to the density
maps $n_0^{Noise}$ and $n_1^{Noise}$ (see Fig. \ref{fig:noise}
(b)), we subtract them to the density maps $n_0$ and $n_1$ we have
previously computed with the phase map of interferogram in Fig.
\ref{fig:C1} b) (in which no noise were introduced). The resulting
error maps $\delta n_0 = n_0 - n_0^{Noise}$ and $\delta n_1 = n_1
- n_1^{Noise}$ are finally confronted. Since the isotropic
component of the density map $n_0^{Noise}$ coincides with the
standard Abel inversion of the symmetrized phase map, we can  to
compare the noise content in the standard and in the generalized
inversions by simply comparing the noise in $n_0^{Noise}$ and
$n_1^{Noise}$. In Fig. \ref{fig:noise}(c) line outs of $\delta
n_0$ and $\delta n_1$ are reproduced.

 Since the density maps are
obtained integrating the uncorrelated  noise with a kernel
$1/\sqrt{\zeta^2-r^2}$ which is rising in approaching  the
symmetry axis ($r\rightarrow 0$), we expect a noise sequence with
a stronger component near $r = 0$. Futhermore, in computing $n_1$
(see Eq. \ref{GenAbel}) we face with the derivative of
$\left({\delta\phi_a(\zeta,x)\over \zeta}\right)$, so that the
$1/\zeta$ term will contribute to enhance the noise in the
$r\rightarrow 0$ region. Nevertheless, because of the
regularization induced by the overall multiplication by $r$, a
balancing of the two effects occur and the noise observed onto the
$n_1^{Noise}$ map is comparable with the one of the standard Abel
inversion $n_0^{Noise}$ map, as it is clear in Fig.
\ref{fig:noise}(c).

To conclude,  the generalized Abel inversion method we propose is
very simple and effective, {\it it uses consistently  the
information carried by the whole phase-shift map} and, as shown in
the examples, it can be successfully applied to asymmetric cases
for which the standard method based on only half-space phase shift
map, fails.

\section*{Acknowledgements}
On of the authors (P.T.) wish to acknowledge support from the
italian M.U.R.S.T. (Project: "Metodologie e diagnostiche per
materiali e ambiente"). Authors are very grateful to D. Giulietti,
from the Dep. of Physics, Univ.  of Pisa and to L.A. Gizzi and R.
Numico from IFAM-CNR, Pisa, for useful discussions and their
encouragement.

\newpage
\section*{Figures Caption}
\begin{figure}[ht]
\caption{ a) Formation of an interferogram. The symmetry axis is
$x$ and the phase-shift is obtained integrating over the $y$
direction. b) Because of an integration along the $y$ axis, no
departure of  $n_e$ from a mirror-symmetric distribution can be
detected. }\label{fig:figura1}
\end{figure}
\begin{figure}[ht]
\caption{ a) The $z-x$ plane projection of the simulated
electronic density. The radial profile is of the form $n^{True} =
n_0\, P_0+n_1\, P_1+n_2\, P_2$, while the density is exponentially
decreasing in the longitudinal direction. b) The simulated
interferogram obtained with the density map $n^{True}$ and
$\lambda_p = 1\, \mu m$. c) and d)  Projections  onto the $z-x$
plane of the density maps obtained with the standard and
generalized Abel inversion of the phase-shift of the simulated
interferogram b). \label{fig:P1}}
\end{figure}
\begin{figure}[ht]
\caption{ a) Line outs of the $z-x$ projection of the true density
profile (dotted line), the sum of the $P_0$ and $P_1$ terms
(dashed line) and of density profiles obtained via standard Abel
inversion of $\delta\phi_s$ (the mean of $\delta\phi^+$ and
$\delta\phi^-$) and of the generalized Abel inversion. The
generalized Abel inversion is considerably more accurate than
standard inversion. b) Line outs of the $z-x$ projection of the
density profiles obtained via standard Abel inversion of
$\delta\phi^+$ and $\delta\phi^-$. None of them well reproduces
the true density profile (dotted line). }\label{fig:P2}
\end{figure}
\begin{figure}[ht]
\caption{ a) The $z-x$ plane projection of the simulated
electronic density. The radial profile is the sum of a background
and a channel not aligned with its symmetry axis, while the
density is exponentially decreasing in the longitudinal direction.
b) The simulated interferogram obtained with the density map
$n^{True}$ and $\lambda_p = 1\, \mu m$. c) Projection onto the
$z-x$ plane of the density maps obtained with the standard Abel
inversion of $\delta\phi^+$  and d) projection of generalized Abel
inversion. \label{fig:C1}}
\end{figure}
\begin{figure}[ht]
\caption{ a) Line outs of the $z-x$ projection of the true density
profile (dotted line) and of density profiles obtained via
standard Abel inversion of $\delta\phi_s$ (dashed line) and of the
generalized Abel inversion (continuous line). The generalized Abel
inversion is considerably more accurate than standard inversion.
b) Line outs of the $z-x$ projection of the density profiles
obtained via standard Abel inversion of $\delta\phi^+$ (dashed
line) and $\delta\phi^-$ (continuous line). None of them well
reproduces the true density profile (dotted line). }\label{fig:C2}
\end{figure}

\begin{figure}[ht]
\caption{ a) Line out of the noisy phase-shift map, obtained by
summing up the phase-shift map of the interferogram in Fig.
\ref{fig:C1} and a Gaussian white noise map. b) Line outs of the
resulting density maps $n_0^{Noise}$ and $n_1^{Noise}$ obtained
via generalized Abel inversion. Line outs of the $n_0$ and $n_1$
maps generated by inverting the noise free map are reported as a
reference. c) Line outs of $\delta n_0$ and $\delta n_1$. The {\it
rms} of the two noise sequences is comparable so the noise content
of $n_0^{Noise}$ and $n_1^{Noise}$ is similar. } \label{fig:noise}
\end{figure}
\newpage
\newpage
\section*{.}
\begin{figure}[ht]
\includegraphics[angle=90,width=10cm]{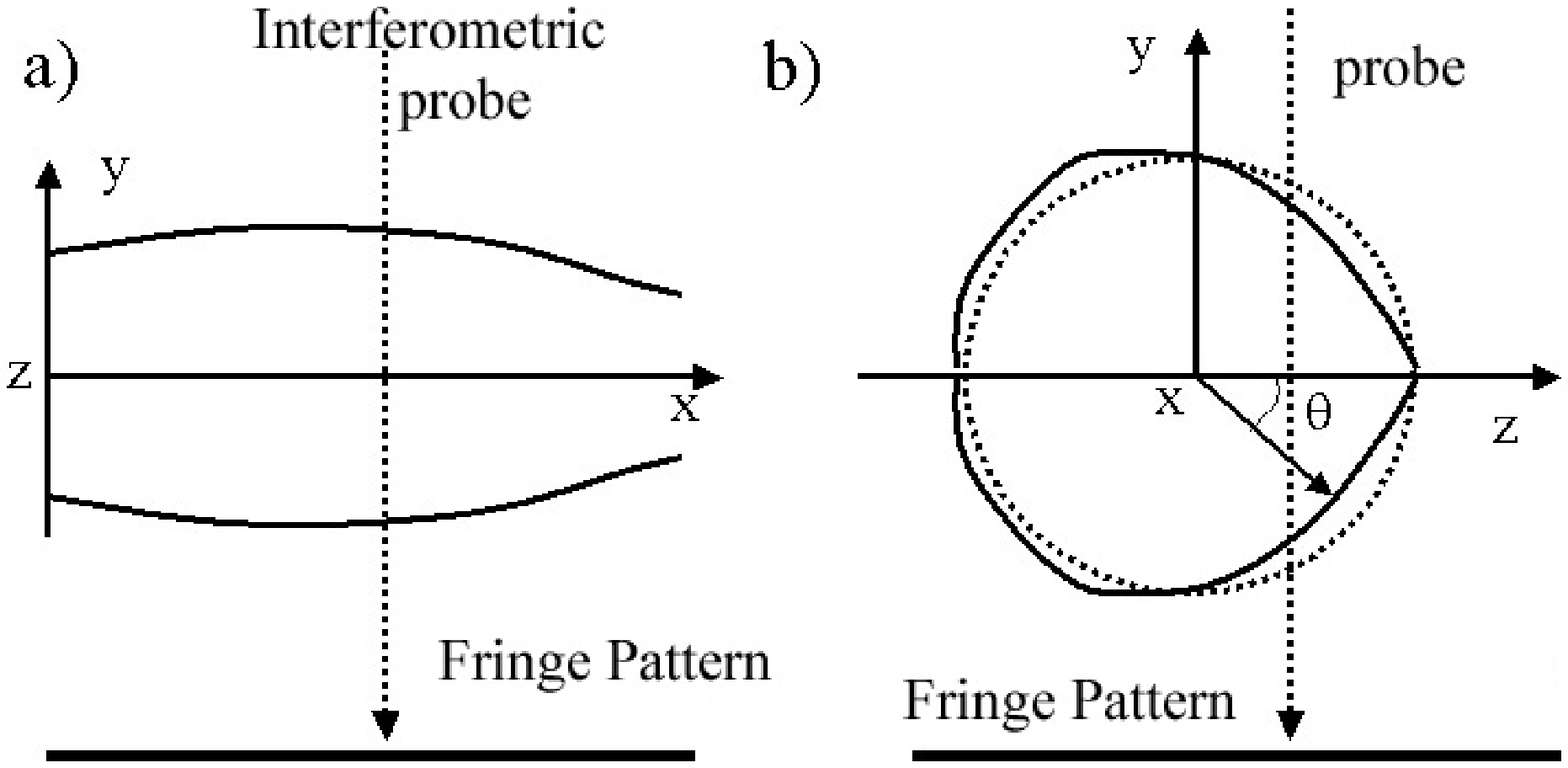}
\end{figure}
\newpage
\begin{figure}[ht]
\includegraphics[angle=90,width=16cm]{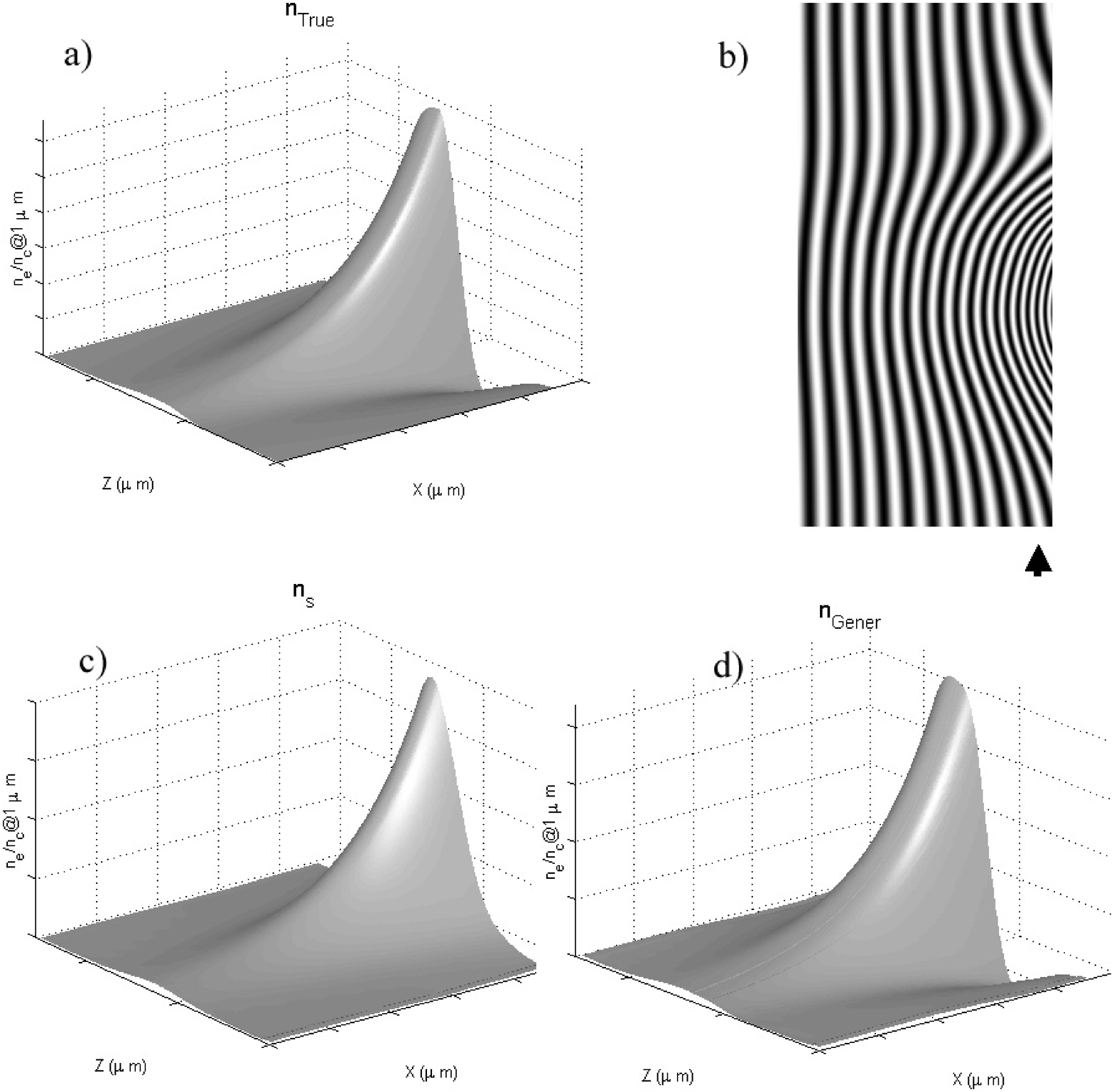}
\end{figure}
\newpage
\begin{figure}[ht]
\includegraphics[angle=90,width=12cm]{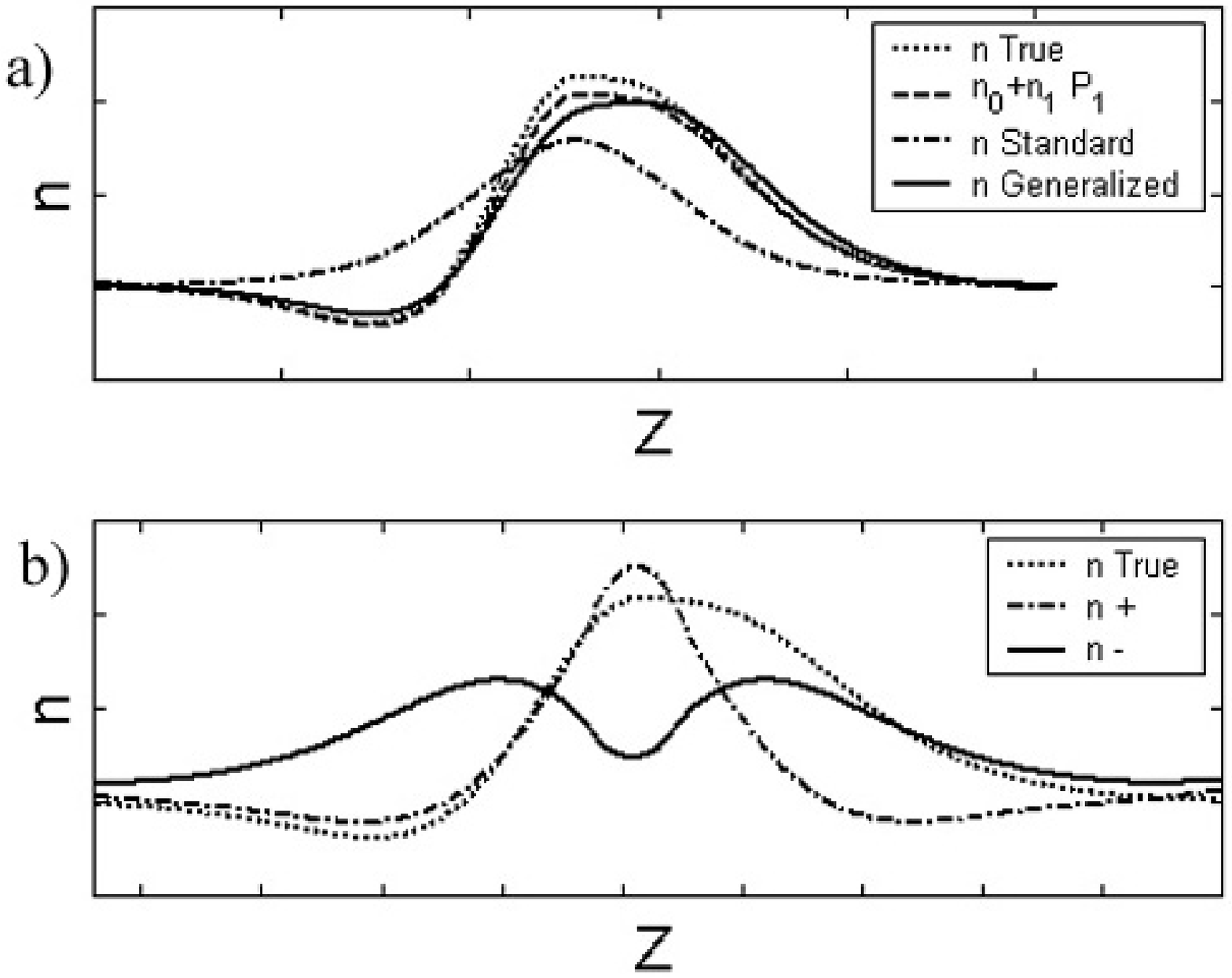}
\end{figure}
\newpage
\begin{figure}[ht]
\includegraphics[angle=90,width=16cm]{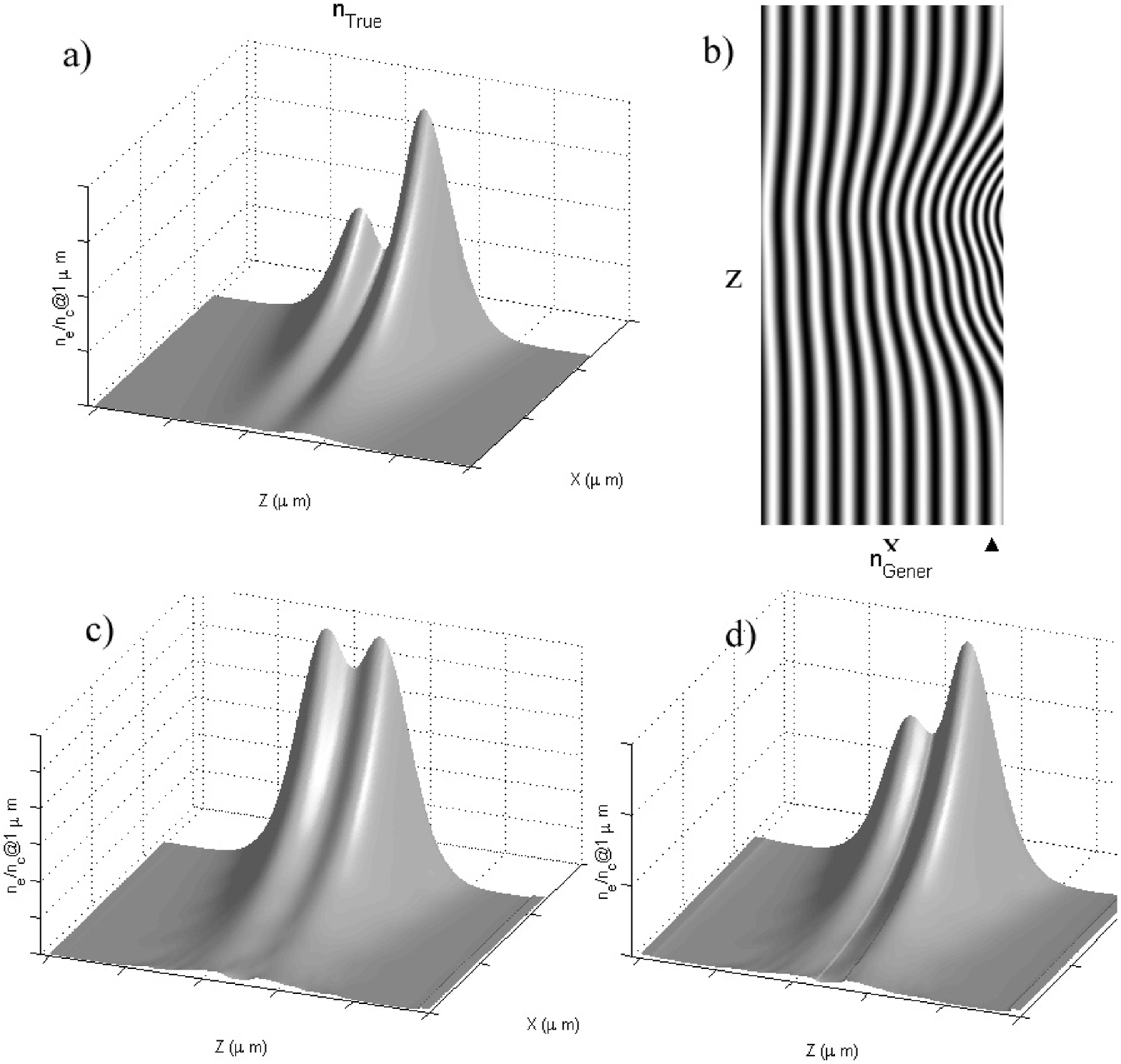}
\end{figure}
\newpage
\begin{figure}[ht]
\includegraphics[angle=90,width=12cm]{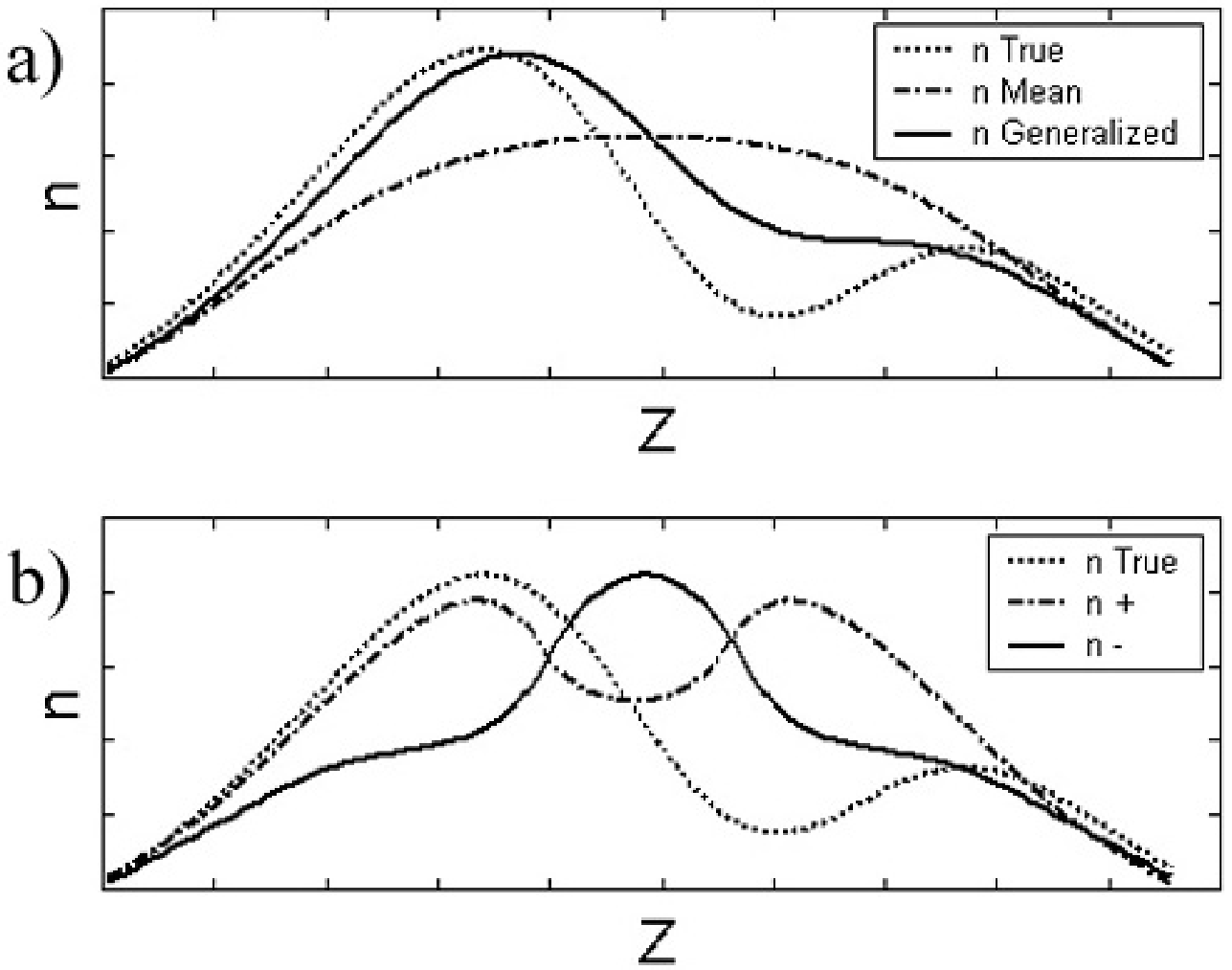}
\end{figure}
\begin{figure}[ht]
\includegraphics[angle=90,width=12cm]{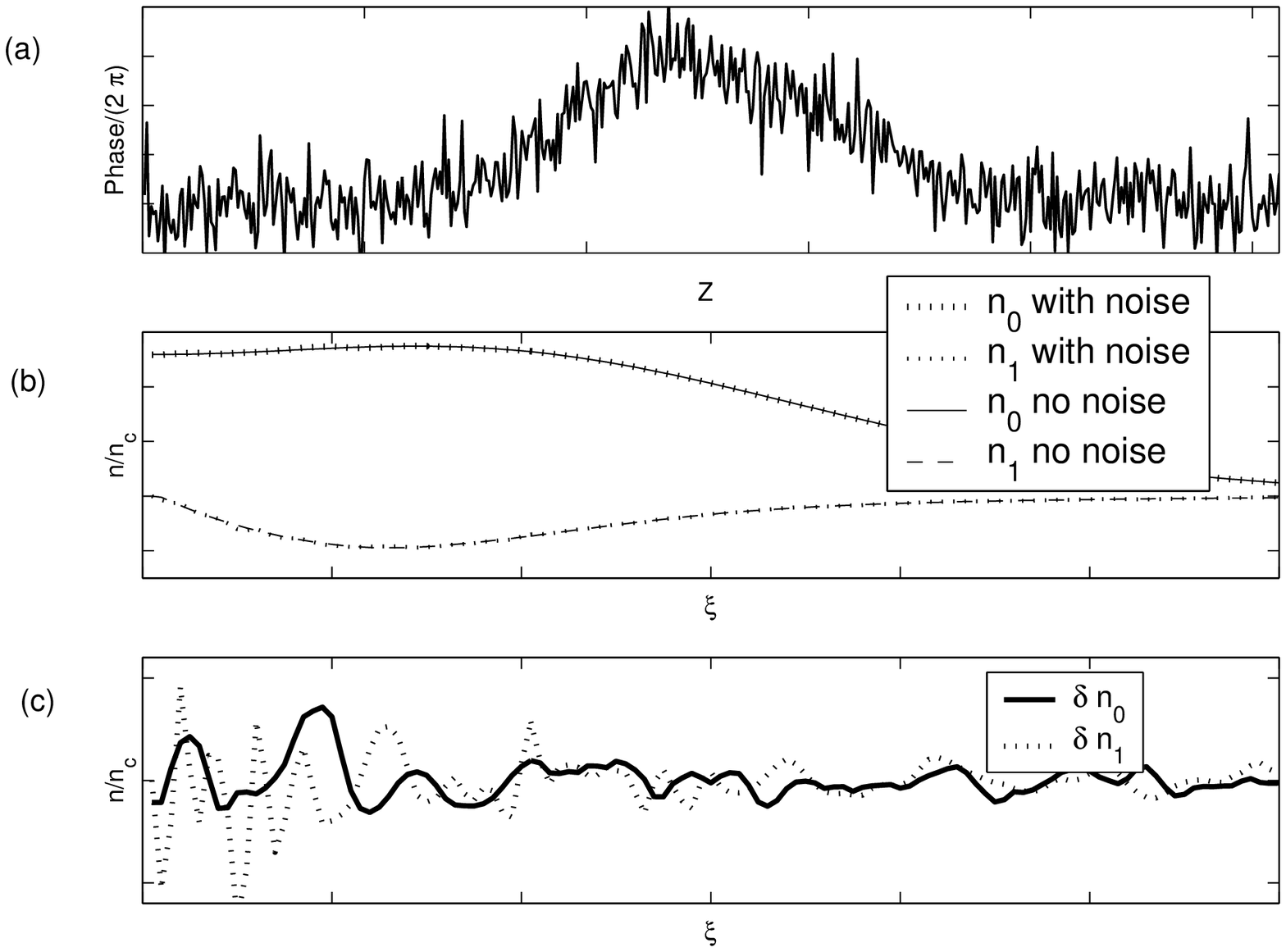}
\end{figure}
\newpage

\end{document}